\documentclass{article}
\usepackage{wrapfig}
\usepackage{amssymb,latexsym,amsmath,color,mathrsfs,ifsym,graphics,stmaryrd}
\usepackage{amsthm, amsfonts, mathrsfs}
\usepackage[colorlinks,linkcolor=blue,urlcolor=blue,citecolor=black,
plainpages=false,pdfpagelabels,breaklinks]{hyperref}
\usepackage[all]{xy}

\usepackage{graphicx}
\usepackage[margin=2.3 cm]{geometry}
\usepackage{caption}

\DeclareMathOperator{\EA}{EA}

\begin{document}

\title{{\sc Multi-Screen Entanglement in\\Tensorial Quantum Mechanics}}

\author{{\sc Christian de Ronde}$^{1,2,3}$, {\sc Raimundo Fern\'andez Mouj\'an}$^{2,6}$, {\sc C\'esar Massri}$^{4,5}$}
\date{}

\maketitle
\begin{center}
\begin{small}
1. Philosophy Institute Dr. A. Korn, University of Buenos Aires - CONICET\\
2. Center Leo Apostel for Interdisciplinary Studies\\Foundations of the Exact Sciences - Vrije Universiteit Brussel\\
3. Institute of Engineering - National University Arturo Jauretche\\
4. Institute of Mathematical Investigations Luis A. Santal\'o, UBA - CONICET\\
5. University CAECE\\
6. Philosophy Institute, Diego Portales University - Santiago de Chile
\end{small}
\end{center}

\begin{abstract}
\noindent In this work we present an invariant-objective formalization of {\it multi-screen entanglement} grounded on Tensorial Quantum Mechanics (TQM) \cite{deRondeFMMassri24c}. This new tensorial formulation of the theory of quanta ---basically, an extension of Heisenberg's matrix mechanics--- allows not only to escape the many problems present in the current account of {\it multi-partite entanglement} grounded on the Dirac-Von Neumann Standard formulation of Quantum Mechanics (SQM) but, more importantly, to consistently represent entanglement phenomena when considering a multiplicity of different screens and detectors.  
\end{abstract}
\begin{small}

\end{small}

\newtheorem{theorem}{Theorem}[section]
\newtheorem{definition}[theorem]{Definition}
\newtheorem{lemma}[theorem]{Lemma}
\newtheorem{proposition}[theorem]{Proposition}
\newtheorem{corollary}[theorem]{Corollary}
\newtheorem{remark}[theorem]{Remark}
\newtheorem{example}[theorem]{Example}
\newtheorem{notation}[theorem]{Notation}

\bigskip

\bigskip

\bigskip

\bigskip

\section*{Multi-Partite Entanglement: A Dead End?}

As remarked by Ingemar Bengtsson and Karol Zyczkowski \cite{BZ} when considering within Standard Quantum Mechanics (SQM) the question: ``Is there any huge qualitative difference between quantum entanglement in composite systems containing three or more subsystems and the known case of bipartite systems? The answer is `yes'.'' Indeed, for bipartite systems while a product state of a pure state $|\psi_1 \rangle \otimes |\psi_2 \rangle$ is considered separable, any other pure state is deemed entangled. These concepts can be naturally generalized for multipartite systems. A state of a system consisting of three subsystems is {\it fully separable} if it has a product form containing three factors, $|\psi_1 \rangle \otimes |\psi_2 \rangle \otimes |\psi_3 \rangle$. All other states are entangled. This appears to be a simple and rather innocent extension, but there is a significant qualitative difference between quantum entanglement in composite systems containing three or more subsystems and the known case of bipartite systems. As explained by Alexander Streltsov \cite{Streltsov23}: ``While for two parties a large number of results in this direction is available, the multipartite setting still remains a major challenge. [...] In contrast to the well-established findings in bipartite systems, the multipartite setting presents a substantially more intricate landscape. Even when considering pure states of three qubits, the current understanding is characterized by isolated results [10-15].'' As revealed by a dimension counting argument, fully separable states are indeed very rare, albeit possessing strange properties. An examination of the mathematical literature shows that ---contrary to the case of matrices where we have, for example, the Schmidt decomposition--- there is no way to find a {\it normal form} (i.e., a representative in each equivalent class) for tensors. Consequently, given two tensors there is no way to know if they both correspond to the same equivalent class. If one considers the number of parties in a quantum composite system, then three is significantly more than two, four is much more than three, etc. Thus, it has become a common place within the specialized literature that, even though fundamental for its development, multi-partite entanglement remains a very delicate and problematic issue within the mainstream research in quantum foundations, quantum information, quantum computation, etc.

As it is well known, in the bipartite setting, one of the main ways to characterize the notion of entanglement is in terms of Local Operations and Classical Communication (LOCC), particularly in relation to transformations involving pure states. Given any pair of pure states $|\psi \rangle$ and $|\phi \rangle$, shared between Alice and Bob, one can try to verify  whether the transformation $|\psi \rangle \rightarrow |\phi \rangle$ is feasible under LOCC. In the asymptotic regime (where many copies of $|\psi \rangle$ are available) we have precise knowledge of the transformation rates, which are intricately linked to the entanglement entropies of the involved quantum states (see \cite{Streltsov23}). But in contrast to the findings in bipartite systems, the multipartite setting presents a substantially more complex landscape and even when considering pure states of three qubits, the current research is characterized by isolated results. Given two arbitrary three-qubit states $|\psi\rangle$ and $|\phi\rangle$, the current knowledge neither allows us to conclusively verify whether $|\psi \rangle$ can be transformed into $|\phi \rangle$ via LOCC with unit probability, nor permits us to determine the optimal asymptotic transformation rate for such a conversion. 

In this work we attempt to argue that these serious problems as well as the complete lack of general results within the research about multi-partite entanglement might be showing us a limit to the mainstream program to quantum entanglement. Taking as a standpoint Tensorial Quantum Mechanics (TQM) \cite{deRondeFMMassri24c}, we will present a new original approach to {\it multi-screen entanglement} with which we will attempt to overcome the many mathematical, conceptual and ---even--- experimental problems that can be found within the research of ``multi-partite entanglement''. The paper is organized as follows. In section 1 we review the many inconsistencies present within the orthodox approach to quantum entanglement. In section 2 we provide a presentation of TQM as an invariant-intensive approach capable to account, through the basis invariance theorem and the factorization invariance theorem, for quantum phenomena in a consistent and coherent manner. Section 3 provides an analysis of quantum entanglement as a multi-screen intensive phenomena. We end, in section 4, with a comparison of the multi-partite approach and the tensorial multi-screen approach.

\section{The Many Inconsistencies of Orthodox Entanglement}

The fact that multi-partite entanglement has been confronted to so many problems and drawbacks when being theoretically and technically developed can be easily explained in terms of the complete lack of a consistent and coherent, systematic foundation of the notion of entanglement itself. As discussed in detail in \cite{deRondeFMMassri24b}, the orthodox definition that is found within the literature in terms of {\it the non-separability of pure states} presented in terms of ``the correlation between microscopic particles'' is complete nonsense. Not only the notions of `purity', `separability' and `particle' have countless problems in the context of QM, but are also incapable to produce any meaningful representation that would allow to explain what is really going on within entanglement phenomena. As it has been shown explicitly in \cite{deRondeMassri21b, deRondeMassri24}, what is referred to as a {\it pure state} within the orthodox literature is an essentially inconsistent notion which confuses several self-contradictory definitions. Sometimes purity is related to an operational result constrained by (binary) certainty (definition 1.1), others it is linked to a completely abstract and purely mathematical account of vectors (definition 1.2) and in other cases it is simply understood as directly linked to the state (an empirical result) after a measurement was actually performed (definition 1.3). 

\begin{definition}[Operational Purity]\label{Opure} Given a quantum system in the state $|\psi \rangle$, there exists an experimental situation linked to that basis (in which the vector is written as a single term) in which the test of it will yield with certainty (probability = 1) its related outcome.\footnote{Von Neumann's application of this notion in the context of quantum logic is also explicit as related to his definition of {\it actual property}, something applied in the many operational approaches that were developed during the 1960s and 1970s (see for a detailed analysis \cite{dDFInternet}). In short, a property is {\it actual} if given a specific experimental set up we know with certainty (probability = 1) the result of the future outcome (see also \cite{Aerts81, Piron76}).} 
\end{definition}

\begin{definition}[Abstract Purity]\label{Apure} An abstract unit vector (with no reference to any basis) in Hilbert space, $\Psi$, is a pure state. In terms of density operators $\rho$ is a pure state if it is a projector, namely, if Tr$(\rho^2) = 1$ or $\rho = \rho^2$. 
\end{definition}

\begin{definition}[Empirical Purity]\label{Epure} A single measurement outcome represented by a one term {\it ket}, $| x_k \rangle$, which, previous to measurement, was originally part of a quantum superposition state $\sum_{i=1}^{N}  c_i | x_i \rangle$.   
\end{definition}

\noindent The confusion between these definitions\footnote{As examples of mainstream textbooks where these different definitions are presented as equivalent see: \cite{Aaronson13, BZ, Mermin07, NC10, Wilde13}.} which are neither equivalent or even consistent has generated a widespread self-contradictory discourse which, of course, can only lead to dead ends. And the same happens with the notion of {\it separability} (of systems into subsystems) which is confused with the mathematical notion of {\it factorizability} (of vectorial spaces). As it was shown in \cite{deRondeFMMassri24b}, this equivalence is completely misleading. While the {\it separability} of systems into subsystems must be understood in terms of classical logic, the mathematical notion of {\it factorizability} implies the {\it projection} to subspaces, something that in mathematical terms is not defined as a separation but as {\it linear surjection}, and that can be intuitively understood as a shadow. Last but not least ---as also discussed explicitly in \cite{deRondeFMMassri24b}---, the widespread discourse applied by mainstream researchers who explain entanglement in terms of the correlation between microscopic ``quantum particles'' is also completely meaningless, lacking any theoretical or experimental support.

Yet, apart from this already flawed foundation, we also find in the literature a fragmentation of the notion of entanglement, which has produced a series of a non-equivalent inconsistent definitions. While some of these new definitions are given in terms of Bell inequalities and different conditions characterizing classical information transfer (such as LOCC, LOSR, etc.), others are mathematically produced in terms of purely abstract geometrical properties (see for a detailed discussion \cite{deRondeFMMassri24b}). As a result, the different sub-fields of research do not even agree about the very basic referent that is being investigated.  This is the reason why, when asking the question: ``What is entanglement?'', to a specialist in the field like Dagmar Bru$\ss$ \cite{Bruss02}, the reply is the following: ``There are many possible answers, maybe as many as there are researchers in this field.'' 

It should come then with no surprise that the extension of entanglement to more than two particles ---what is known in the filed as ``multi-partite entanglement''--- has become an intractable problem. A dead end which exposes the natural consequence of a completely inconsistent foundation. However, an escape route of this maze has been, in fact, proposed. Following the thread of invariance and objectivity ---namely, the basic preconditions of rational scientific discourse--- the logos categorical approach to QM has developed a consistent and coherent formal-conceptual representation of quantum phenomena. This proposal has been able to reconfigure an operational-invariant formalism that is able to account for a state of affairs independently of specific viewpoints, measurement situations or reference frames ---thus escaping contextuality. Going back to Heisenberg's mathematical formalism and focusing on {\it intensive values} ---instead of binary measurement outcomes--- it is indeed possible to bypass the famous Kochen-Specker theorem \cite{deRondeMassri21a} and restore a truly invariant mathematical representation of a situation that remains {\it the same} even when considered from different perspectives or viewpoints ---let them be formal or experimental. In turn, the addition of specifically designed operational concepts capable to address in a qualitative fashion those same phenomena, already described in quantitative terms, has allowed a natural and simple extension of the mathematical formalism itself from matrices to tensors. This is what has been recently presented as Tensorial Quantum Mechanics (TQM) \cite{deRondeFMMassri24c}, a development to which we now turn our attention. In the following sections it will become clear why this new formulation opens the doors to a new formal-conceptual consistent and coherent account of what we call multi-screen entanglement.

\section{Tensorial Quantum Mechanics}

When following the thread of invariance, step by step, it becomes evident which is the reference of the mathematical formalism, which are the {\it moments of unity}, the essential concepts that need to be operationally defined within the theory such as {\it screen}, {\it detector} and {\it experimental arrangement} (see \cite{deRondeFMMassri24a} and \cite{deRondeFMMassri24c}). In turn, this allows also to consistently account for the physical content of {\it bases} and {\it factorizations} ---something problematic within the orthodox formulation (see \cite{deRondeFMMassri24a}). In the following we give a self contained explanation of the formalism of TQM (see for a more complete description of TQM \cite{deRondeFMMassri24c}) as well as a detailed definition of the main physical notions that are required in order to bridge the gap between the mathematical formalism and its conceptual representation.

We begin with the definition of a (simple) \emph{graph} as a pair $G = (V, E)$, where $V$ is a set whose elements are called vertices (or nodes), and $E$ is a set of unordered pairs $\{v,w\}$ of vertices, whose elements are called edges. While each {\it vertex} is related to the mathematical notion of {\it projector operator} and to the physical concept of  {\it power of action}, each {\it edge} is linked to the mathematical concept of {\it commutation} and the {\it experimental compatibility} of powers within the same measurement set-up. 

\begin{definition} {\bf Graph of powers:} Given a Hilbert space $H$, the graph of powers $G(H)$ is defined such that the vertices are the projectors on $H$ (called \emph{powers}), and an edge exists between projectors $P_1$ and $P_2$ if they commute.
\end{definition} 

\noindent It is these powers, in their multiplicity and their relationships, that allow us to define an \emph{Intensive State of Affairs} (ISA) ---in contrast to the binary {\it Actual State of Affairs} (ASA) that is used to represent situations in classical physics and relativity. But first, we need to formalize the notion of intensity (or potentia). The assignment of intensities is called {\it Global Intensive Valuation} (GIV).

\begin{definition} {\bf Global Intensive Valuation:}  A Global Intensive Valuation is a map from $G(H)$ to the interval $[0,1]$.
\end{definition}

\noindent Clearly, not all GIVs are compatible or consistent with the relations between powers. We will focus on those that define an ISA as follows:

\begin{definition} {\bf Intensive State of Affairs:} Let $H$ be a Hilbert space of infinite dimension. An \emph{Intensive State of Affairs} is a GIV $\Psi: G(H)\to[0,1]$ from the graph of powers $G(H)$
such that $\Psi(I)=1$ and 
\[
\Psi(\sum_{i=1}^{\infty} P_i)=\sum_{i=1}^\infty \Psi(P_i)
\]
for any piecewise orthogonal operator $\{P_i\}_{i=1}^{\infty}$. The numbers $\Psi(P) \in [0,1]$ are called {\it intensities} or {\it potentia} and the vertices $P$ are called \emph{powers of action}. Taking into consideration the ISAs, it is then possible to advance towards a consistent GIV which can bypass the contextuality expressed by the Kochen-Specker Theorem \cite{deRondeMassri21a, KS}. 
\end{definition} 

\begin{definition} {\bf Quantum Laboratory:} We use the term \emph{quantum laboratory} (or quantum lab or Q-Lab) as the operational concept of ISA. 
\end{definition} 

\noindent Within a Q-Lab, we have the concepts of screen, detector and experimental arrangement as well as their specific relation to factorizations and bases. 

\begin{definition}{\bf Screen and Detector:} A \emph{screen} with $n$ places for $n$ detectors corresponds to the vector space $\mathbb{C}^n$. Choosing a basis, say $\{|1\rangle,\dots,|n\rangle\}$, is the same as choosing a specific set of $n$ {\it detectors}. A \emph{factorization} $\mathbb{C}^{i_1}\otimes\dots \otimes\mathbb{C}^{i_n}$ is the specific number $n$ of screens, where the screen number $k$ has $i_k$ places for detectors, $k=1,\dots,n$. Choosing a \emph{basis} in each factor corresponds to choosing the specific detectors; for instance $|\uparrow\rangle, |\downarrow\rangle$. After choosing  a basis in each factor, we get a basis of the factorization $\mathbb{C}^{i_1}\otimes\dots \otimes\mathbb{C}^{i_n}$
that we denote as
\[
\{ |k_1\dots k_n\rangle \}_{1\le k_j\le i_j}.
\]
\end{definition} 

\begin{definition}{\bf Power of action:} The \emph{ basis element} $|k_1\dots k_n\rangle$ determines the \emph{ projector}  $|k_1\dots k_n\rangle \langle k_1\dots k_n|$ which is the formal-invariant counterpart of the objective physical concept called \emph{ power of action} (or simply \emph{power}) that produces a global effect in the $k_1$ detector of the screen $1$,  in the $k_2$ detector of the screen $2$ and so on until the $k_n$ detector of the screen $n$. Let us stress the fact that this effectuation does not allow an explanation in terms of particles within classical space and time. Instead, this is explained as a characteristic feature of powers. In general, any given power will produce an intensive multi-screen non-local effect. 
\end{definition} 

\begin{definition} {\bf Experimental Arrangement:} Given an ISA, $\Psi$, a factorization $\mathbb{C}^{i_1}\otimes\dots \otimes\mathbb{C}^{i_n}$ and a basis $B=\{|k_1\dots k_n\rangle\}$ of cardinality $N=i_1\dots i_n$, we define an \emph{ experimental arrangement} denoted $\EA_{\Psi,B}^{N,i_1\dots i_n}$, as a specific choice of screens with detectors together with the potentia of each power, that is,
\[
\EA_{\Psi,B}^{N,i_1\dots i_n}= \sum_{k_1,k_1'=1}^{i_1}\dots \sum_{k_n,k_n'=1}^{i_n} 
\alpha_{k_1,\dots,k_n}^{k_1',\dots,k_n'}|k_1\dots k_n\rangle\langle k_1'\dots k_n'|.
\]
\end{definition} 
\begin{definition}{\bf Potentia:} The number that accompanies the power $|k_1\dots k_n\rangle \langle k_1\dots k_n|$ is its \emph{ potentia} (or intensity) and the basis $B$ determines the powers defined by the specific choice of screens and detectors. The number $N$ which is the cardinal of $B$ is called the \emph{ degree of complexity} (or simply degree) of the experimental arrangement. 
\end{definition} 

Assume now that in a Q-Lab we want to change or modify an experimental setup. We have two theorems that allow us to predict the possible outcomes that will obtained in the new experimental arrangement. If the number of powers (i.e., the degree of complexity) remains the same after the rearrangement, then the {\it Basis Invariance Theorem} tell us that the new experimental arrangement is equivalent to the previous one, but if the complexity of the new experimental arrangement drops, then the {\it Factorization Invariance Theorem} tell us that all the knowledge in the new experimental arrangement was already contained in the previous one (see for a detailed analysis \cite{deRondeFMMassri24a, deRondeMassri23}). 

\begin{theorem}{\sc (Basis Invariance Theorem)}
Given a specific QLab $\Psi$, all experimental arrangements of the same complexity are equivalent independent of the basis. 
\end{theorem}

\begin{theorem} {\sc (Factorization Invariance Theorem)}
The experiments performed within a $\EA_{\Psi}^N$ can also be performed with an experimental arrangement of higher complexity N+M, $\EA_{\Psi}^{N+M}$, that can be produced within the same QLab $\Psi$.  
\end{theorem}

In order to treat formally this situation, we must work with tensors. Specifically, assume that we have two bases $B$ and $B'$ obtained from two experimental arrangements in the same Q-Lab $\Psi$,
\[
B = \{|k_1\dots k_n\rangle\}_{1\le k_j\le i_j}, \quad
B' = \{|\kappa_1\dots \kappa_{m}\rangle\}_{1\le \kappa_j\le \iota_j}, \quad
\]
From $B$ we infer that the first experimental arrangement has $n$ screens, where the first screen has $i_1$ detectors, the second $i_2$ detectors and so on. The second experimental arrangement has $m$ screens, where the first screen has $\iota_1$ detectors, the second 
$\iota_2$ and so on. Assume that the two bases are related by the following transformation,
\[
|k_1\dots k_n\rangle = \sum_{\kappa_1,\dots,\kappa_m=1} ^{\iota_1,\dots,\iota_m}
\lambda_{\kappa_1,\dots,\kappa_m}^{k_1\dots k_n}
|\kappa_1\dots \kappa_{m}\rangle
,\quad 1\le k_1\le i_1,\dots,1\le k_n\le i_n.
\]
Then, we can convert the first experimental arrangement into the second one through the algebraic properties of the tensors. If
\[
\EA_{\Psi,B} = \sum_{k_1,k_1'=1}^{i_1}\dots \sum_{k_n,k_n'=1}^{i_n} 
\alpha_{k_1,\dots,k_n}^{k_1',\dots,k_n'}|k_1\dots k_n\rangle\langle k_1'\dots k_n'|
\]
then,
\[
\EA_{\Psi,B'} = 
 \sum_{\kappa_1,\dots,\kappa_m=1} ^{\iota_1,\dots,\iota_m}
 \sum_{\kappa_1',\dots,\kappa_m'=1} ^{\iota_1,\dots,\iota_m}
\left(
\sum_{k_1,k_1'=1}^{i_1}\dots \sum_{k_n,k_n'=1}^{i_n} 
\alpha_{k_1,\dots,k_n}^{k_1',\dots,k_n'}
\lambda_{\kappa_1,\dots,\kappa_m}^{k_1\dots k_n}
\overline{\lambda_{\kappa_1',\dots,\kappa_m'}^{k_1'\dots k_n'}}
\right)
|\kappa_1\dots \kappa_{m}\rangle
\langle\kappa_1'\dots \kappa_{m}'|.
\]
Notice that in standard multi-index notation (a notation more suited for these type of algebraic expressions) the last equation can be written more compactly as
\[
\EA_{\Psi,B'} = \sum_{\kappa,\kappa'}
\left(\sum_{k,k'}
\alpha_{k}^{k'}
\lambda_{\kappa}^{k}
\overline{\lambda_{\kappa'}^{k'}}
\right)
|\kappa\rangle\langle\kappa'|.
\]
We can then summarize all these definitions in the following table showing the strict relation between the conceptual and mathematical levels of representations.
\begin{center}
\begin{tabular}{|l|l|}
\hline
Physical concepts&Mathematical concepts\\
\hline
Q-Lab $\Psi$ & ISA $\Psi:G(H)\to[0,1]$\\
Screen with $n$ places &   $\mathbb{C}^n$\\
$n$ detectors &    Basis $\{|1\rangle,\dots,|n\rangle\}$\\
$n$ screens where screen $j$ has $i_j$ places &    Factorization  $\mathbb{C}^{i_1}\otimes\dots \otimes\mathbb{C}^{i_n}$  \\
Detectors in each screen &  $ \{ |k_1\dots k_n\rangle \}_{1\le k_j\le i_j}$ \\
Power of action & Projector $|k_1\dots k_n\rangle \langle k_1\dots k_n|$ \\
A Q-Lab $\Psi$ with screens and detectors& The tensor $\EA_{\Psi,B}^{N,i_1\dots i_n}$\\
Degree of complexity $N$ & Cardinal of $B$\\
Potentia of the power  &
The real number $\alpha_{k_1,\dots,k_n}^{k_1,\dots,k_n}$\\
Modify an experimental arrangement&Change of basis\\
\hline
\end{tabular}
\captionof{table}{Relations between physical and mathematical concepts.}
\end{center}

Notice that while the case of a single screen corresponds to the orthodox vectorial approach, the case of two screens is linked to the orthodox extension to density matrices. It is important to remark that in our tensional formulation these are particular cases of the more general situation where we have $n$ screens. These leads to what we call a ``multi-screen analysis of entanglement'' that will attempt to replace the multi-partite approach that is discussed within the literature. In the following sections we discuss this proposed extension.

\section{Entanglement as a Multi-Screen Intensive Phenomena}

In the framework of TQM, and contrary to SQM, it becomes thus completely straight forward to work with any number of screens and detectors. Adding screens or detectors does not generate an essential complication or difficulty and, furthermore, there is a very natural way in which the different EAs can be graphically represented. Let us start with the case of one single screen with one detector,
\begin{center}
\includegraphics[scale=.3]{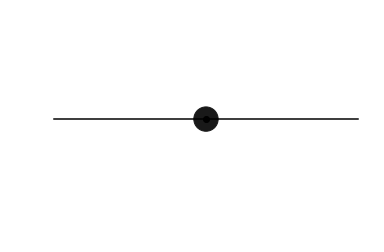}
\captionof{figure}{{\small One screen with one detector
where\\ the only power of action has intensity 1.}}
\end{center}
If we go to the case of two detectors in one screen we recover the known situation of a Stern-Gerlach experiment described by two powers with a specific potentia each. 
\begin{center}
\includegraphics[scale=.3]{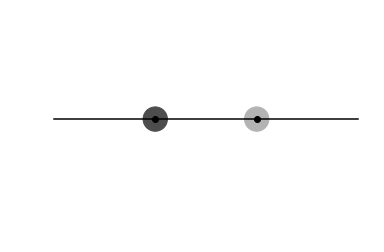}
\captionsetup{justification=centering}
\captionof{figure}{{\small One screen with two detectors where the first power of action\\ has intensity 0.7 and the second power has intensity 0.3.}}
\end{center}
Notice that the following representation is what the orthodox literature considers a {\it pure state} $|k\rangle$ in a six dimensional Hilbert space obtainable with certainty (see \cite{deRondeMassri21b}). 
\begin{center}
\includegraphics[scale=.3]{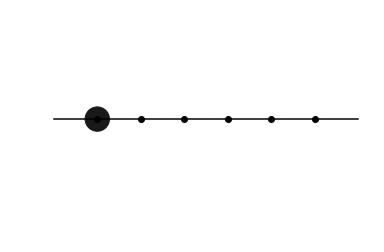}
\captionsetup{justification=centering}
\captionof{figure}{{\small One screen with six detectors where\\ the first power has intensity 1.}}
\end{center}
It is easy to see why, contrary to SQM, where the notion of pure sate is kernel and contra-posed to the notion of {\it mixture}, in TQM this is the less interesting case, the one with less complexity, and consequently, the one that provides less knowledge with respect to the state of affairs in question. While in the case of one screen powers are represented by points in the screen, in the case of two screens powers can be represented by lines. 
\begin{center}
\includegraphics[scale=.45]{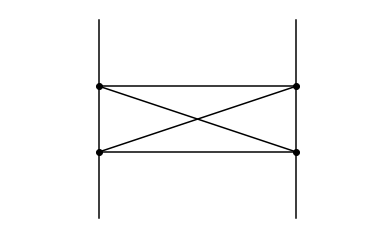}
\includegraphics[scale=.45]{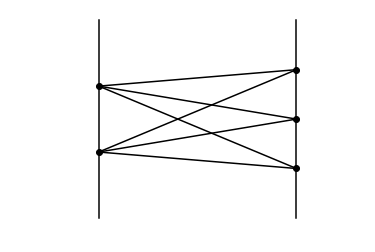}\\
\includegraphics[scale=.45]{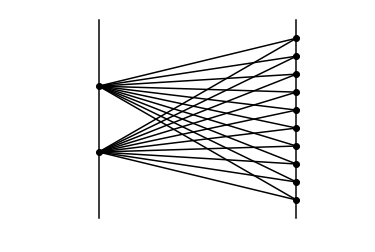}
\includegraphics[scale=.45]{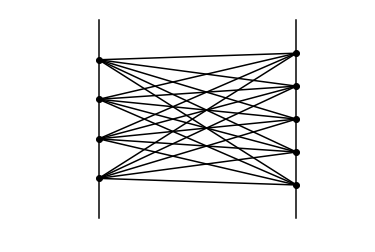}
\captionof{figure}{{\small Two screens with two different configurations of detectors.}}
\end{center}
If we add one more screen the powers become filled triangles like we show in the next two examples
\begin{center}
\includegraphics[scale=.45]{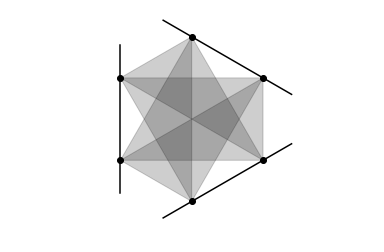}
\includegraphics[scale=.45]{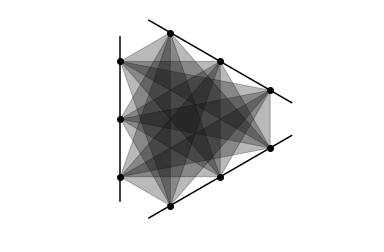}
\captionsetup{justification=centering}
\captionof{figure}{{\small Three screens with two and three detectors where\\ each different power is depicted as a specific triangle.}}
\end{center}
In order to produce a more \emph{readable} figure, it may be convenient to depict only some of the powers. For example,
\begin{center}
\includegraphics[scale=.45]{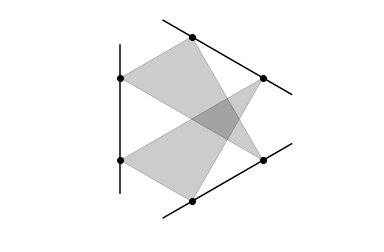}
\includegraphics[scale=.45]{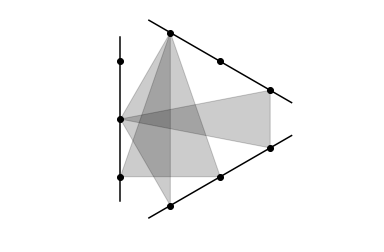}
\captionof{figure}{{\small Three screens with two detectors each.\\Some of the powers are shown.}}
\end{center}
Notice that we can add several screens without entering in a contradiction with the set of concepts that we already developed. In particular, for $n$ screens the powers will be represented by filled polytopes of dimension $n$. For example, 
\begin{center}
\includegraphics[scale=.45]{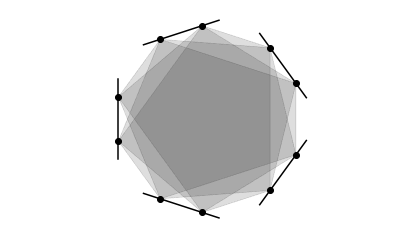}
\includegraphics[scale=.45]{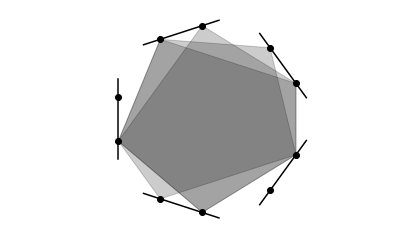}
\captionof{figure}{{\small Five screens with two detectors each.\\In the left, all the powers are shown.}}
\end{center}
\begin{center}
\includegraphics[scale=.45]{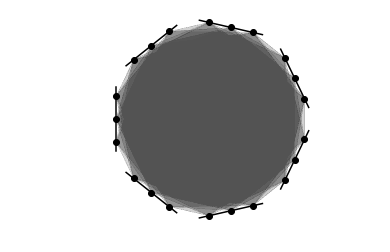}
\includegraphics[scale=.45]{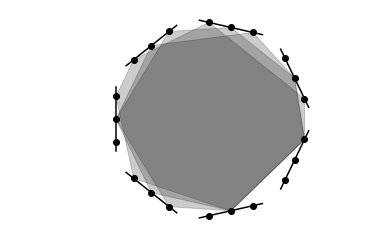}
\captionof{figure}{{\small Seven screens with three detectors each.\\In the left, all powers are represented.}}
\end{center}

\section{Multi-Partite Entanglement vs Multi-Screen Entanglement}

There are many advantages to the multi-screen proposal when compared to the multi-partite approach. First, while in the multi-partite case the problem of more than three screens becomes intractable, within the multi-screen approach we can work with any number of screens and detectors without any complication. Contrary to the multi-partite approach, given an EA of a certain complexity it allow us to produce  through the {\it Factorization Invariance Theorem} any EA of leaser complexity. And also, while in the multi-partite approach entanglement is basis dependent \cite{DelaTorre10, Earman15}, within the multi-screen approach the {\it Basis Invariance Theorem}, allows to compute the precise intensities in any chosen basis. Another strength of the multi-screen approach, contrary to the multi-partite scheme, is that it allows us to work with $n$ screens and $m$ detectors with the same basic notions used within the simplest cases of one or two screens with few detectors.

It is interesting to notice that, even though there are some fundamental conceptual differences, our approach shares a few similarities with that of Deutsch and Hayden in terms of {\it descriptors} \cite{Bedard20}, and those of Raymond-Robichaud in terms of {\it noumenal states} \cite{RR17} and {\it evolution matrices} \cite{RR20}. All these approaches, as demonstrated by Bedard, are equivalent \cite{Bedard21}. In \cite{DH00}, Deutsch and Hayden  demonstrated that a complete description of a composite quantum system can be deduced from the complete descriptions of its subsystems, setting aside the concept of non-locality. The key component of these approaches is to consider, say, a {\it qubit} not merely as something with two degrees of freedom, but as a system with infinitely many degrees of freedom. The first similarity regards the use of Heisenberg's matrix mechanics as a fundamental mathematical tool, more powerful than the orthodox Dirac-von Neumann vectorial approach. In this respect, it is important to remark that our proposal goes even further by developing a tensorial formulation of QM which, in turn, allows us to develop our multi-screen entanglement approach and consider as many detectors and screens as desired. The second similarity concerns the need to increase the number of {\it degrees of freedom} in order to provide a complete description of the state of affairs (see \cite{Bedard21}). According to our proposal, the greater the complexity of an experimental arrangement, the more knowledge we obtain of that state of affairs (see \cite[Sec. 5]{deRondeFMMassri24b}). This idea, of course, goes against the mainstream focus on {\it pure states} which, in our terms, have {\it minimal complexity} and, consequently, provide the least possible amount of knowledge of a given state of affairs.

Multipartite entanglement plays a crucial role in various quantum information processing tasks, such as quantum teleportation and dense coding  (see  for a detailed discussion \cite{beng2016, shang2023, neumann2008, xie2024}). However, as it is well known, this approach presents serious obstacles:
\begin{itemize}
\item Algorithmic Complexity: describing the structure of quantum entanglement for multipartite systems is problematic due to the infinite number of entanglement classes, even for a simple four-qubit system.
\item Quantification: while there are various measures designed for bipartite systems, such as concurrence, negativity, and entanglement of formation, quantifying entanglement for multipartite systems is much more complex. This is because a single Schmidt coefficient governs the two-qubit entanglement, making all the entanglement monotones for the two-qubit system equivalent.
\item Genuine Requirement: most proposed measures do not meet the ``genuine requirement'', which states that a genuine multipartite entanglement measure should only, and always, vanish for biseparable states. This is crucial to accurately quantify the collective strength of the parties entanglement and unlock their potential in quantum tasks.
\item Geometric Complexity: extending measures like the triangle measure for three-qubit systems to systems with more than three qubits is problematic due to the over-flexibility of geometric shapes.
\item Optimization Procedures: evaluating measures like relative entropy and some other 
geometric measures requires using optimization procedures which are highly complicated, even in the case of {\it pure states}.
\end{itemize}
These obstacles make the study of multipartite entanglement a complex field full of difficulties and dead ends which add to the long list of problems already present within SQM itself. However, applying instead TQM, all these obstacles and problems, as shown in \cite{deRondeFMMassri24b, deRondeFMMassri24c}, simply disappear. 

Let us end with an example that illustrates the simplicity and benefits of our approach. Assume we want to study what happens if we remove one screen from an experimental arrangement with four screens with two detectors each.
\begin{center}
\includegraphics[scale=.7]{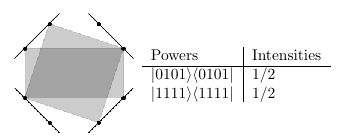}
\captionof{figure}{{\small Four screens with two detectors where \\ 2 different powers are depicted as a specific squares.}}
\end{center}
In this case, the experimental arrangement is given by the following tensor:
\[
\EA_{\Psi,B} = \frac{1}{2}|0101\rangle\langle0101|+\frac{1}{2}|1111\rangle\langle1111| .
\]
Now, if we remove the fourth screen, that is, if we consider the transformation
\[
\mathbb{C}^2\otimes\mathbb{C}^2\otimes\mathbb{C}^2\otimes\mathbb{C}^2\to\mathbb{C}^2\otimes\mathbb{C}^2\otimes\mathbb{C}^2,\quad
|k_1k_2k_3k_4\rangle\to|k_1k_2k_3\rangle,
\]
the resulting experimental arrangement has three screens with two detectors.
\begin{center}
\includegraphics[scale=.7]{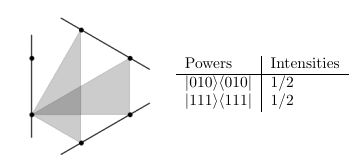}
\captionof{figure}{{\small The new experimental arrangement after removing a screen.}}
\end{center}
The new experimental arrangement is then given by the following tensor:
\[
\EA_{\Psi,B'} = \frac{1}{2}|010\rangle\langle010|+\frac{1}{2}|111\rangle\langle111| .
\]

\section*{Acknowledgements} 

The authors state that there is no conflict of interest. This work was partially supported by the following grants: ANID-FONDECYT, Project number: 3240436. We want to thank Alan Forrester for recommending us pertinent bibliography.

\end{document}